# Thermal Stability and Topological Charge Fragmentation in Antiskyrmions of Rhombohedral Barium Titanate

Florian Mayer[1*] and Jiří Hlinka[2]

[1] Materials Center Leoben Forschung GmbH, Roseggerstrasse 12, 8700 Leoben, Austria

[2] Institute of Physics of the Czech Academy of Sciences, 182 00, Praha 8, Czech Republic

*florian.mayer@mcl.at

## Abstract

Antiskyrmions, as topological quasi-particles, hold significant promise for spintronics and nanoscale data storage applications. Using molecular dynamics simulations based on effective Hamiltonians, we investigate the thermal stability of antiskyrmion nanodomains in rhombohedral barium titanate. At 1 K, antiskyrmions with a topological charge of -2 emerge as the most stable nanodomain state across all examined diameters. In our systematic study, the most robust antiskyrmion was found to have a diameter of 4 nm, maintaining its original size, shape, and topological charge up to the characteristic temperature $T^* \approx 85$ K. Domains with diameters between 2.8 and approximately 4.5 nm exhibit fragmentation into six topological defects, termed quarks, each carrying a fractional skyrmion charge of -1/3. For domains larger than 4.5 nm, each topological quark splits into two pre-quarks, each with a charge of -1/6. These larger nanodomains demonstrate increased mobility and a growing tendency for shape and skyrmion charge fluctuations. Above the $T^*$ temperature, all larger nanodomains gradually shrink to a diameter of about 4 nm before collapsing into a single-domain state. These findings reveal the relatively high stability of antiskyrmions over a broad temperature range, even in the absence of a stabilizing bias field, and emphasize the pivotal role of topological quark dynamics. This establishes barium titanate as a key platform for exploring and applying topological phenomena.

## I.    INTRODUCTION

The discovery of topological defects in various materials has brought significant advances to the field of condensed matter physics [1–3], revealing new insights and offering promising technological applications. Among these phenomena, magnetic skyrmions, nontrivial swirling spin textures, have transformed the understanding of nanoscale magnetism and paved the way for innovations in spintronics [4–8]. Motivated by these magnetic skyrmions, recent research has turned to exploring analogous topological objects in ferroelectric materials [9,10]. In ferroelectrics, the coupling of electric polarization with other degrees of freedom, such as strain and lattice distortions, can lead to the formation of topological defects. While much of the initial focus was on ferroelectric skyrmions in superlattices [9,11–13], theoretical work has since revealed a broader class of topological defects in ferroelectric systems, including the discovery of ferroelectric antiskyrmions in bulk rhombohedral barium titanate (BaTiO$_3$) by Gonçalves et al. [14]. These topological defects in electric polarization textures, with a negative geometrically invariant topological charge, represent an intriguing class of nanodomains with hexagonal-like shapes and intricate polarization patterns stabilized by the anisotropy of polarization correlations [14]. The overall net topological charge of -2 for this antiskyrmion was found to arise from six line defects, referred to as quarks, each carrying topological charge of -1/3. More recently, these findings were corroborated by phase-field simulations based on the generalized Ginzburg-Landau-Devonshire model, which focused on the mechanisms of formation, transport and annihilation of antiskyrmions [15]. In parallel, a 2D Ginzburg-Landau study inspected whether the unwrapped skyrmions, also called skyrme lines, can form in the rhombohedral ferroelectric barium titanate [16]. These unwrapped skyrmions can be considered as 180-degree ferroelectric domain walls in a 3D material, containing particular line defects with a fractional topological charge [16]. The antiskyrmions and unwrapped skyrmions in the rhombohedral barium titanate substantially extend the already rich variety of topological defects that can emerge in ferroelectric materials [17–20], and it motivates further exploration of the interplay between topology, lattice effects, and polarization textures.

Building on these studies, this work investigates the thermal stability and behavior of induced nanodomains of varying sizes in the rhombohedral barium titanate. It is clear from previous studies [14,15] that, at zero bias field, the creation of isolated antiskyrmions increases the system's energy. Consequently, thermal fluctuations can eventually overcome the protective potential barriers, leading to a decay into the homogeneous single domain state. Understanding the effect of thermal fluctuations on these topological defects is thus crucial for both potential technological applications and the design of the experimental validation of these phenomena. In contrast to the approach by Gonçalves et al. [14], which employed an atomistic shell potential [21], this study utilizes effective Hamiltonians for atomistic simulations [22–25], offering improved computational efficiency [26,27] and enhanced accuracy in predicting transition temperatures [28]. Molecular dynamics (MD) simulations based on effective Hamiltonians enable a systematic study across a wide range of nanodomain sizes while employing a reasonably detailed heating protocol. This approach aims to explore more thoroughly the evolution and thermal stability of ferroelectric antiskyrmions as a function of temperature.

Initially, the study investigates nanodomains at 1 K, confirming that antiskyrmions [14] with a topological charge of -2 represent the most stable skyrmionic nanodomain state across all induced scenarios. Nanodomains with diameters smaller than approximately 4.5 nm exhibit a highly symmetric configuration of six evenly distributed topological quarks around the domain perimeter, each carrying a fractional charge of -1/3 [14]. For larger diameters, the nanodomain ground state retains its overall 3m point group symmetry and stability. However, topological quarks are observed to stretch and split into pairs of novel defects with charges of -1/6, denoted here as pre-quarks, employing the term used for hypothetic objects considered in the elementary particle theory [29]. The study then systematically explores the impact of temperature on the antiskyrmion structure, revealing that nanodomains with diameter larger than 4.5 nm remain quite stable up to approximately 100 K, despite structural adaptations and increasing topological quark mobility. However, at temperatures exceeding the characteristic temperature $T^* \approx 85$ K, all large nanodomains eventually shrink to a critical diameter of about 4 nm before collapsing into a single-domain state.

The structure of this paper is as follows: Section II details the methodology and simulation settings. Section III.A discusses the formation of antiskyrmions at 1 K. Section III.B provides an analysis of local charge distribution and division of topological quarks. Section III.C explores the threshold temperature and the evolution of nanodomain diameters as a function of temperature. Section III.D investigates the evolution of topological charge across temperatures. Section III.E explores the antiskyrmion configurations preceding their thermally activated collapse.

## II. COMPUTATIONAL DETAILS

This study utilizes MD simulations based on the effective Hamiltonian approach [22,23,30], involving the careful selection of a phonon subspace, local basis establishment [31,32], and parametrization through Density Functional Theory (DFT) calculations [22]. The applied Hamiltonian, as illustrated in Equation 1, involving various interactions, is specifically tailored with an extended parameterization from Ref. [28], known for its incorporation of additional anharmonic couplings [28,33] to higher energy phonons, resulting in improved agreement with experimental transition temperatures [28,34]. In essence, the applied Hamiltonian comprises several energy contributions. The kinetic energies of the optical local soft-mode $\boldsymbol{u}$ and the acoustic mode $\boldsymbol{w}$, the local-mode self-energy $V^{self}$, dipole-dipole interaction $V^{dpl}$, short range interaction $V^{short}$, elastic energies $V^{elas}$, strain-phonon interactions $V^{coupl}$ and interaction with an external field. The elastic energies as well as the strain-phonon coupling are further divided into homogeneous and inhomogeneous parts. The inhomogeneous parts are directly related to the acoustic displacement $\boldsymbol{w}$ in the long-wavelength limit and provide information about the inhomogeneous deformation of the local unit cells. The homogeneous components indicate the deformation of the entire supercell associated with the strain variables. The time derivatives of the variables are denoted by $\dot{u}_\alpha$ and $\dot{w}_\alpha$, with $\alpha$ indicating the Cartesian component. $\boldsymbol{R}$ denotes a translation vector indicating the position of the unit cells within the supercell. $\eta$ represents strain variables in Voigt notation. $Z^*$ denotes the Born effective charge associated with the soft mode $\boldsymbol{u}$. The braces { } indicate a set of amplitudes $\boldsymbol{u}$ and $\boldsymbol{w}$ within a supercell. $M^*_{dipole}$ and $M^*_{acoustic}$ are the effective masses of the soft mode and acoustic mode, respectively. The external field $\epsilon$ is incorporated through its vector product with each local dipole vector $Z^*\boldsymbol{u}$.

$$\begin{aligned} H_{BTO}^{eff} = &\frac{M^*_{dipole}}{2}\sum_{R,\alpha} \dot{u}_\alpha^2(\boldsymbol{R}) \\ &+ \frac{M^*_{acoustic}}{2}\sum_{R,\alpha} \dot{w}_\alpha^2(\boldsymbol{R}) + V^{self}(\{\boldsymbol{u}\}) + V^{dpl}(\{\boldsymbol{u}\}) + V^{short}(\{\boldsymbol{u}\}) \\ &+ V^{elas,homo}(\eta_1,\dots,\eta_6) + V^{elas,inho}(\{\boldsymbol{w}\}) + V^{coupl,homo}(\{\boldsymbol{u}\},\eta_1,\dots,\eta_6) \\ &+ V^{coupl,inho}(\{\boldsymbol{u}\},\{\boldsymbol{w}\}) - Z^*\sum_R \epsilon \cdot u(\boldsymbol{R}) \end{aligned}$$

(1)

All MD simulations presented in this work utilized the freely available software, *feram* [24,35,36]. The canonical ensemble was employed, with temperature control facilitated by a velocity scaling thermostat, validated against Nose-Poincaré [37] results. The supercell size for all simulations discussed in this work was standardized to 64x64x64. A finite time step of 2 fs was utilized for the simulations. The total simulation times were rigorously checked for convergence, and the chosen values, based on thermalization conducted over 800 ps with subsequent averaging over a 200 ps interval, ensure accurate results. In the heating runs, the concluding configuration of the preceding temperature served as the initial configuration for the subsequent temperature. The heating runs were initiated at 1 K and conducted in increments of 5 K up to a temperature of 211 K. The initial setup for the MD simulation at a temperature of 1 K, which also serves as the starting point for subsequent heating runs, closely follows the methodology outlined in the work by Gonçalves et al. [14]. Specifically, a rhombohedral phase (R3m) [38,39] was adopted, aligning the polarization direction along the <111> crystallographic axis. To achieve this, the local amplitudes of the effective Hamiltonian were straightforwardly oriented in the <111> direction. Subsequently, a nearly cylindrical nanodomain was induced along the <111> direction, including an inversion of the polarization direction within this direction. The resulting configuration, encapsulating the specified nanodomain characteristics, was then employed as the starting point for the MD simulation. It is crucial to note that solely the initial configuration for the first-time step at 1 K was supplied to the MD simulation. No additional boundary conditions (except the periodic boundary conditions) or fixations were imposed thereafter, affording the simulation complete freedom within the potential energy surface. This approach ensures that the system evolves dynamically, reflecting the intrinsic behavior dictated by the underlying forces and interactions. In total, 19 different starting configurations, spanning diameters from 2.2 nm to 13 nm at 0.6 nm intervals, were generated and subsequently simulated.

The associated topological charge $Q$ was calculated using the formulation proposed by Berg and Lüscher [40], as illustrated in the work of Heo et al. [41]. In the latter work, the calculation was proposed as evident in Equations 2 and 3. Here, the summation is carried out over the areas $A_l$ of the spherical triangles spanned by the spin vectors $\boldsymbol{m}_i$, $\boldsymbol{m}_j$, and $\boldsymbol{m}_k$. The sign is determined by the relation $sign(A_l) = sign[\boldsymbol{m}_i \cdot (\boldsymbol{m}_j \times \boldsymbol{m}_k)]$.

$$Q = \frac{1}{4\pi}\sum_l A_l$$

$$\cos\left(\frac{A_l}{2}\right) = \frac{1 + \boldsymbol{m}_i \cdot \boldsymbol{m}_j + \boldsymbol{m}_j \cdot \boldsymbol{m}_k + \boldsymbol{m}_k \cdot \boldsymbol{m}_i}{\sqrt{2(1 + \boldsymbol{m}_i \cdot \boldsymbol{m}_j)(1 + \boldsymbol{m}_j \cdot \boldsymbol{m}_k)(1 + \boldsymbol{m}_k \cdot \boldsymbol{m}_i)}}$$

(2)

(3)

An advantage of effective Hamiltonians is the direct access they provide to local dipoles, as these are intimately linked to the local mode amplitudes employed in the simulation as described above. Furthermore, the selection of elementary triangles must align with the prescribed definition found in Ref. [41]. For an accurate assessment of the topological charge, it is essential to utilize the normalized vectors corresponding to the local dipole moments.

## III. RESULTS

### A. Low-Temperature Antiskyrmions

To investigate the emergence of topological structures at low temperatures, simulations were performed at 1 K using supercells generated as described in Section II, spanning a range of nanodomain diameters. For nanodomains with diameters below 2.2 nm, no distinct topological structures were observed. Instead, the dipoles within the domains merged seamlessly with the surrounding matrix, creating a uniformly polarized supercell along the <111> direction and resulting in a single-domain state.

Figures 1a and 1b illustrate the dipole configurations for nanodomains with diameters of 2.8 nm and 4.0 nm, respectively. The plots represent projections of three consecutive (111) planes along the <111> direction, providing a visualization of the dipole arrangements. To standardize the analysis, rotated axes (x′ and y′) were defined such that the <111> direction aligns with the z-axis. The color map in the figure differentiates upward (+z) from downward (-z) dipoles, while arrows illustrate the local dipole directions. Importantly, these plots depict local polarization vectors, which are proportional to the local dipole moments and vary only by the volume of the corresponding region. Both nanodomains shown in Figure 1a and 1b exhibit a hexagonal-like dipole arrangement in the (111)-plane, characterized by the 3m point group symmetry. The arrangement features six vortices evenly distributed around the perimeter of the domain, with an equal number of clockwise and counterclockwise rotations. This pattern results from the relaxation of dipoles around the domain, forming a stable and energetically favorable configuration. Furthermore, these configurations confirm the skyrmionic ground-state symmetry of Goncalves et al. [14], with each domain exhibiting six evenly distributed topological quarks carrying a fractional charge of -1/3, as discussed in detail in the next section. The calculated net topological charge of -2 for both 2.8 nm and 4.0 nm domains is consistent with the previous findings of Gonçalves et al. [14], confirming the presence of antiskyrmions in these small domains. As the diameter is further increased, slight deviations from the highly symmetric structure become apparent. However, the overall 3m point group symmetry stays intact. For instance, Figure 1c illustrates a minor distortion characterized by elongation in the y' direction, yet the topological charge is still -2. With additional increases in diameter, various other asymmetric metastable states are observed. However, the most common configurations are shown in Figures 1d through 1h, where the 3m point group symmetry is preserved and the topological charge remains at -2. These structures thus appear to represent the skyrmionic nanodomain ground state configuration across a wide range of diameters.

This study at 1 K confirms that all nanodomains from 2.8 nm to 13 nm maintain the topological ground state symmetry with a topological charge of -2, irrespective of minor variations in symmetry. The observed configurations align with the intrinsic stability of the system.

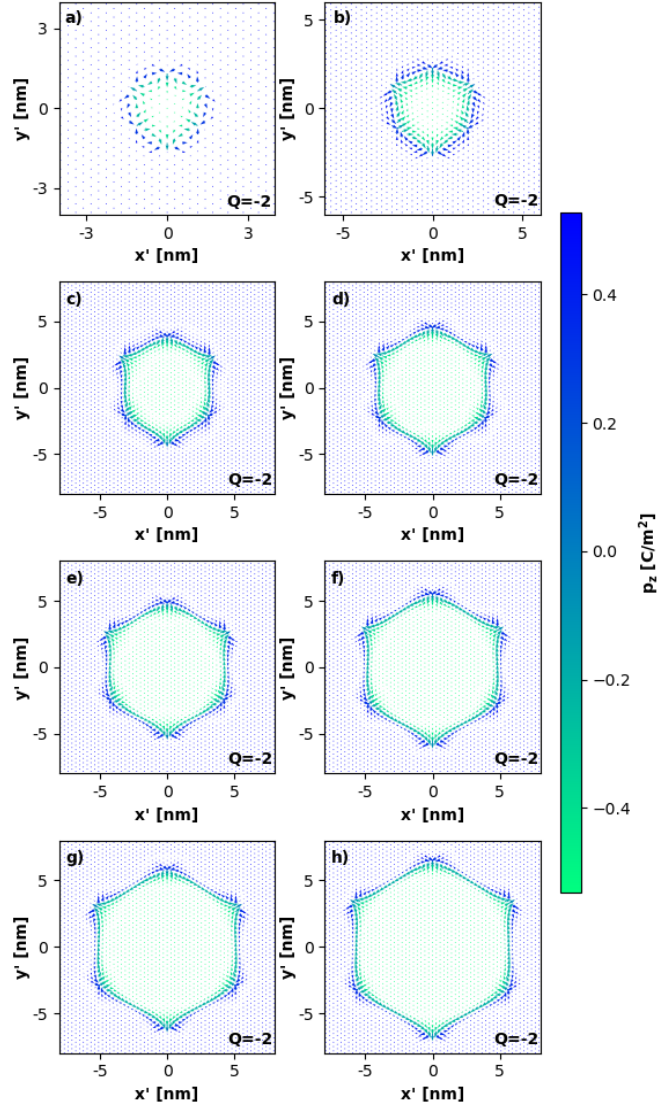

*Figure 1. Emergence of antiskyrmions at a low temperature of 1 K. The plot illustrates a projection of three subsequent cross-sections parallel to the (111)-plane within the 64x64x64 supercell. The arrows indicate the time-averaged (200 ps) local polarization vector, whereas the color represents the out of plane ($p_z$) component of the polarization vector. The diameters of the induced nanodomains are 2.8 nm (a), 4 nm (b), 6.4 nm (c), 8.2 nm (d), 9.4 nm (e), 10.6 nm (f), 11.8 nm (g), and 13 nm (h), respectively. Q represents the topological charge.*

## B. Division of Topological Quarks

In this section, we investigate the size-dependent changes in the antiskyrmion overall shape and fractionalization of skyrmion charge density. To provide context, we summarize the findings of Gonçalves et al. [14], who observed that antiskyrmions in rhombohedral barium titanate exhibit a hexagonal-like shape. This structure results from the crystallographic constraints of the rhombohedral phase, where polarization vectors align preferentially along the <111> directions of the parent cubic structure. This alignment leads to a hexagonal cross-section for the antiskyrmion core, surrounded by in-plane polarization vortices with alternating inward and outward orientations, stabilizing a 3m point group symmetry conform to the rhombohedral phase.

Gonçalves et al. [14] also highlighted the antiskyrmion's unique topological properties, including localized regions with fractional topological charges of approximately -1/3, referred to as topological quarks. These quarks are positioned near the centers of the edges of the hexagonal-like structure, where abrupt changes in polarization occur. The sum of these fractional charges results in a total topological charge of -2. This intricate internal structure distinguishes these antiskyrmions from other topological defects like Bloch or Néel skyrmions [17].

In the following analysis, we delve deeper into the nature of the topological quarks. We begin with Figure 2a, which depicts the smallest antiskyrmion in our study, with a diameter of 2.8 nm. This antiskyrmion exhibits the highly symmetric structure consisting of forming a hexagon-like pattern with six distinct polarization vortices centered at the midpoints between adjacent vertices of this hexagon. To visualize the topological quarks, we plot the corresponding local topological charge density in Figure 2b. The six evenly distributed topological quarks of the dominant negative charge are clearly coinciding with the cores of the polarization vortices, while the minor positive residual charges are located at the vertices of the hexagon.

As established in the previous section, all domain sizes retain the total topological charge of -2. However, what about the distribution of the topological charge density? To address this, we examine an intermediate-sized antiskyrmion with a diameter of 5.8 nm, shown in Figure 2c. This is the antiskyrmion with a slight diameter elongation along the y' axis. The vortices in this direction are slightly larger, involving more local dipoles. The corresponding local charge density is shown in Figure 2d, revealing an intriguing phenomenon: the -1/3 topological quarks are stretched and split. The largest splitting is observed along the two larger edges of the hexagon-like pattern, so that the -1/3 quark divides into two -1/6 topological pre-quarks. Even in the four shorter edges, the splitting is also evident, though not as pronounced. These observations suggest that as the vortices expand, the topological quarks split. Integrating locally over these newly formed pre-quarks confirms a fractional charge of -1/6. This splitting clearly linked to the object's overall size and the corresponding diameter expansion.

This phenomenon becomes even more pronounced in Figure 2e, which shows an antiskyrmion with a diameter of 10.6 nm. Here, the system regains 3m point group symmetry, with all quasi-hexagon edges being uniformly sized. The local charge density in Figure 2f reveals that all -1/3 topological quarks have split, resulting in 12 -1/6 topological pre-quarks instead of the original six -1/3 quarks. Despite this splitting, the total topological charge remains -2. The same behavior is observed in Figures 2g and 2h, where an even larger antiskyrmion is depicted.

In summary, the ground state corresponds to a topological charge of -2 with a highly symmetric structure comprising six vortices. In nanodomains with diameter smaller than 4.5 nm, these vortices can be associated with six -1/3 topological quarks. For larger nanodomains these -1/3 topological quarks split into -1/6 topological pre-quarks. At the same time, the larger nanodomains are more prone to adopt lower symmetry configurations.

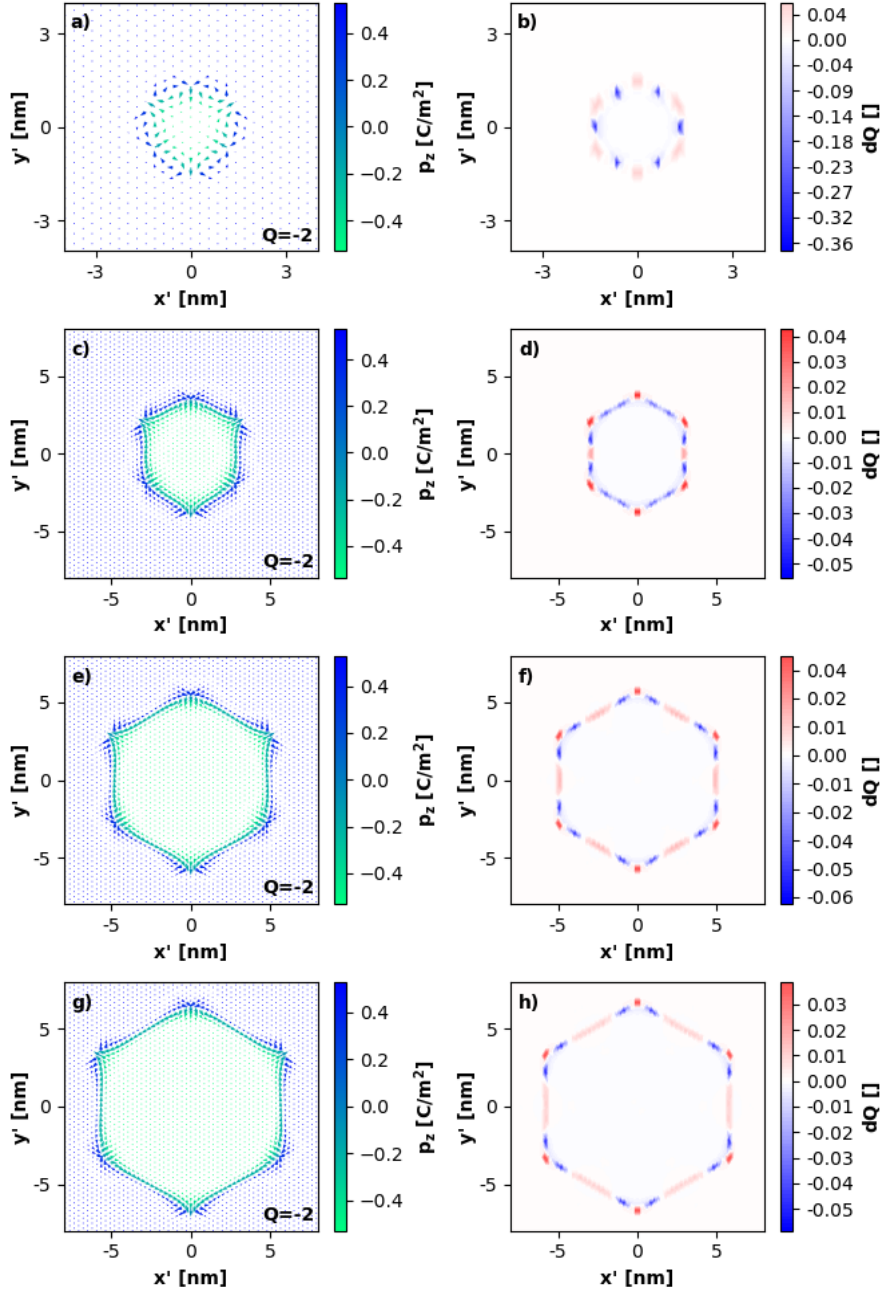

*Figure 2. Selected antiskyrmions and their local topological charge distributions. (a) The highly symmetric ground-state antiskyrmion with a charge of -2. (b) The corresponding local charge distribution with six -1/3 charge quarks. (c) A slightly larger antiskyrmion with a charge of -2. (d) Onset of stretching and splitting of -1/3 quarks. (e) An intermediate antiskyrmion with a charge of -2. (f) Visual splitting of -1/3 quarks into -1/6 pre-quarks. (g) A larger antiskyrmion, confirming the splitting behavior as evident in (h).*

## C. Temperature Evolution of Antiskyrmion Diameter

This section explores the thermal stability of antiskyrmions. Figure 3 illustrates how the diameter of antiskyrmions evolves with temperature. The diameter is defined as the shortest distance between the opposite quasi-hexagon edges on the representative cross-section and measured by tracking the transition points where polarization switches from upward (+z direction) to downward (-z direction). While the shape variation has some impact on the assigned diameter value, this method provides a reasonable and computationally convenient estimate of the antiskyrmion size.

The threshold temperature $T_{thr}$ is defined as the temperature at which nanodomains collapse into a single-domain state, resulting in the disappearance of antiskyrmions. This temperature varies depending on the size of the nanodomains, as evident from Figure 3. Considering the evolution of diameters, the following trend is observed: Small diameters, up to approximately 4.5 nm, exhibit a nearly constant behavior. For example, the smallest antiskyrmion, corresponding to a nanodomain with a diameter of 2.8 nm, maintains a constant size and has a threshold temperature of 16 K. As the diameter increases, the threshold temperature rises significantly. Comparing the evolution of diameters and their corresponding threshold temperatures, the antiskyrmion with a diameter of 4 nm emerges as the most stable. At this size, both the diameter and shape remain constant up to a threshold temperature of 85 K. This associated temperature, will be further referred to as the characteristic temperature $T^* = 85\,K$, and is defined by the configuration's maximum stability against temperature.

When the induced diameter exceeds 4 nm, the antiskyrmions begin to shrink as the temperature increases. The onset of this decrease can be determined by a depinning temperature, $T_m$, which is lower than the characteristic temperature $T^*$. The larger the diameter, the lower the $T_m$. An empirical relationship can be used to describe this behavior, with the depinning temperature roughly expressed as $T_m = 100\text{K}\,(1 - d/d_m)$ where $d_m \approx 14$ nm. Similarly, the threshold temperature $T_{thr}$ can be empirically described as a function of the diameter $d$, based on the results shown in Figure 3. This relationship for the threshold temperature can be, for example, expressed by a function of the form $T_{thr} = 80\text{K} + 12\text{K}\ln[(d - d_0)/d_0]$ and setting $d_0 = 2.8$ nm.

Additional simulations were conducted by directly heating the initial nanodomains equilibrated at 1 K to the corresponding last temperature below effective threshold and the one just above it. These simulations, carried out over 1 ns at these two near-threshold temperature points for each given diameter, confirmed that antiskyrmions could remain stable for 1 ns as long as the temperature remained below this effective threshold temperature. Above this threshold, no antiskyrmions survived, confirming that the threshold temperature represents an abrupt and robust limit of the thermal stability at the corresponding 1 ns timescale.

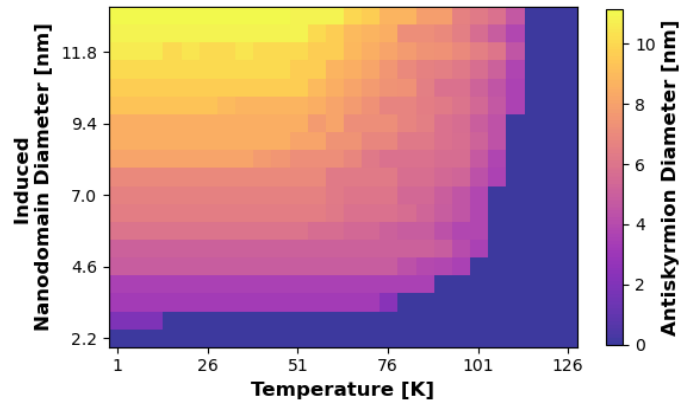

*Figure 3. Temperature evolution of the antiskyrmion diameter for various initial nanodomain diameters induced at 1 K.*

### D. Evolution of Topological Charge

This section examines how the topological charge of induced nanodomains evolves with temperature, building upon the findings from the previous analysis. The topological charge at each temperature was calculated using time-averaged local dipole moments, consistently referencing to the projection of three consecutive (111) planes to ensure precision across all measurements (see Sections II and III.A for methodology). Figure 4 illustrates the change in topological charge as temperature increases, with white regions indicating temperatures at which the nanodomains collapse.

For all induced diameter scenarios at 1 K, a topological charge of -2 is observed. As previously discussed, this provides strong evidence that this state represents the skyrmionic nanodomain ground state. The temperature evolution of the topological charge remains highly stable across a wide temperature range. For the smallest diameters, the -2 state is maintained consistently throughout heating, corresponding to the constant diameters of these structures. For diameters larger than 4.5 nm, however, the total topological charges start to fluctuate above $T_Q = 110\text{K} - \alpha\sqrt{d/d^* - 1}$, $T_Q > T_m$. Here, we set $d^* = 2.8$ nm and $\alpha$ was estimated by fitting the onset of fluctuations in Figure 4, resulting in a value of approximately 36 K. The alternative topological charges appearing in Figure 4 are integer values such as -3, -1, 0, and +1. It means that topological quarks can not only split and move but also acquire positive fractional charges. These exotic states are energetically close to the skyrmionic ground state and can be accessed under the influence of thermal energy.

In summary, the -2 state is by far the most stable configuration, observed consistently across a wide range of temperatures and diameters. Larger induced diameters, however, undergo substantial structural evolution and topological transformations driven by thermal energy.

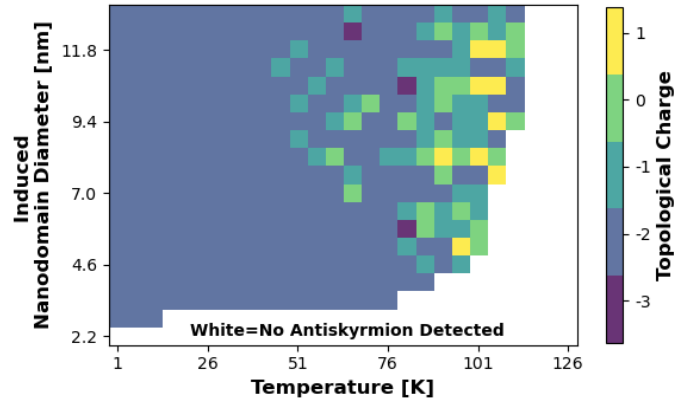

*Figure 4. Evolution of topological charges as a function of induced nanodomain diameter and temperature.*

### E. Thermally Induced Collapse of Large Antiskyrmions

In this final section, we examine the structural changes induced by thermal energy just before reaching the threshold temperature. To this end, trajectories with various initial diameters were compared, and the corresponding cross-sections at temperatures 5-20 K below the critical threshold are shown in Figure 5. As seen in Figure 5, the diameters of all examined trajectories are rather similar, despite their initially induced differences in size. For example, Figure 5h depicts a nanodomain initially induced with a diameter of 12.4 nm. This observation indicates that all nanodomains with the initial diameter larger than 4.5 nm tend to decrease the overall domain wall surface and regardless of the initial diameter the final collapse occurs when the inner diameter of the cross-section through the inverted region reaches about 4 nm. It also suggests that the dependence of the threshold temperature for these large nanodomains on their initial diameter is only a kinetic effect.

In this context, it is worth emphasizing that the nanodomain with an elongated ellipsoidal cross section, as shown in Figures 5b, 5d, and 5f, are farther from collapse than their more isotropic counterparts depicted in Figures 5a, 5c, and 5e. This is evident from the differences in their respective threshold temperatures as well as the remaining simulation times.

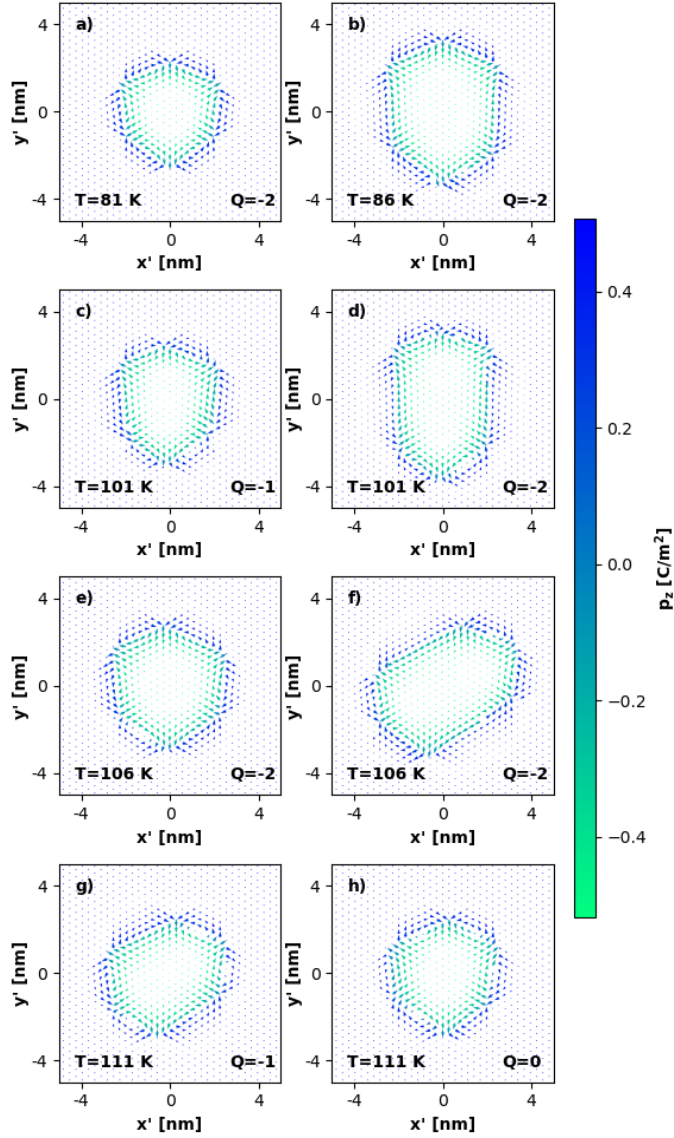

*Figure 5. Visualization of antiskyrmions before disappearance. The arrows indicate the time-averaged (200 ps) local polarization vector, whereas the color represents the out of plane ($p_z$) component of the polarization vector. The initial diameters of the induced nanodomains at 1 K are 4 nm (a), 5.2 nm (b), 6.4 nm (c), 7.6 nm (d), 8.8 nm (e), 10 nm (f), 11.2 nm (g), and 12.4 nm (h), respectively. Q represents the topological charge.*

## IV. FINAL REMARKS AND SUMMARY

This study primarily focused on the size effects and thermal fluctuations influencing antiskyrmions in barium titanate. Notably, there is remarkable quantitative agreement between the 1 K results obtained here and the atomistic shell model molecular dynamics simulations of Gonçalves et al. [14]. Despite being based on fundamentally different methodologies, one relying on interatomic pair potentials and the other on a systematic anharmonic phonon expansion, both approaches identified the smallest stable antiskyrmion as having the same diameter of 2.8 nm.

The characteristic temperature $T^* \approx 85$ K, which defines the stability range of optimal-diameter antiskyrmions, is arguably the most significant result, as it provides valuable insight for future experimental considerations. While this quantity is likely sensitive to the specific details of the model, it is also a plausible and realistic value. This is supported by the fact that the current effective Hamiltonian reproduces the experimental phase transition temperatures in barium titanate with a precision of 10-20 K [28]. In any case, the present results establish an initial benchmark for exploring potential strategies to extend the stability range of antiskyrmions in barium titanate and related materials.

The most intriguing observation of this work is undoubtedly the clear splitting of quarks into pre-quarks, as described and documented in Figure 2. For the largest nanodomains, the skyrmionic topological charge density is nonzero along the circumference of nearly perfect hexagons. The pre-quarks appear as 12 marked blue spots located approximately 1 nm from the vertices (see, for example, Fig. 2h). While there is some residual positive topological charge at the vertices and edges of the hexagon, emphasized in the color coding, the dominant contribution to the overall integer charge arises solely from the 12 blue spots. In other words, large nanodomains generally exhibit topologically trivial 180-degree domain walls, with anomalies confined to the 1 nm vicinity of the vortices. As the nanodomain diameter decreases, the pre-quarks gradually approach each other, and eventually, each pair of pre-quarks merges to form a single quark positioned at the center of the edge connecting two vortices. Interestingly, these configurations with condensed quarks correspond to the most stable nanodomains.

Considering the temperature behavior, at the lowest temperatures, where thermal fluctuations are negligible, the optimal nanodomain geometry consistently exhibited high symmetry and an overall antiskyrmionic topological charge of -2, regardless of the nanodomain diameter. In small antiskyrmions, with diameters between 2.8 and 4 nm, the topological charge was distributed among six distinct quarks, each carrying a fractional topological charge of -1/3. These small antiskyrmions retained their skyrmionic topological charge of -2 and maintained a roughly constant geometry up to their diameter-dependent threshold temperature. In contrast, for antiskyrmions with diameters larger than 4 nm, the topological quarks were split into pairs of -1/6 pre-quarks, which shifted toward the vertices of the hexagonal-like nanodomain cross-section. These larger antiskyrmions proved more susceptible to thermal fluctuations. As the temperature increased, domain walls and topological pre-quarks became mobile, leading to fluctuations in the overall shape and a gradual reduction in diameter. At even higher temperatures, the larger nanodomains underwent more significant structural changes, resulting in fluctuations in their overall topological charge. Above $T^*$, the inner diameter of the nanodomain approached the critical 4 nm limit, eventually undergoing thermal collapse into a single-domain state.

The demonstrated stability of antiskyrmions across a broad temperature range highlights barium titanate as a promising material for investigating and utilizing topological phenomena, both in simulations and experimental studies. This research not only deepens our understanding of topological defects but also opens new avenues for practical applications.

# ACKNOWLEDGMENTS

This work was supported by the Czech Science Foundation (project no. 19-28594X) and by the European Union's Horizon 2020 research and innovation programme under grant agreement no. 964931 (TSAR). The authors gratefully acknowledge Mauro A. P. Gonçalves, Marek Paściak, and Vilma Stepkova from FZU for sharing preliminary results and valuable discussions.